\documentclass[cits]{PoS}

\title{Simulation of the Polarized Sky at 1.4 GHz }

\ShortTitle{Polarized Sky at 1.4 GHz }

\author{\speaker{Shane P. O'Sullivan}$^a$, J. M. Stil$^b$, A. R. Taylor$^b$, R. Ricci$^b$, J. K. Grant$^b$, K. Shorten$^b$\\%
       $^a$Department of Physics, University College Cork, Ireland\\
       $^b$Department of Physics and Astronomy, University of Calgary, Alberta, Canada\\
       E-mail: \email{shaneosullivan@physics.ucc.ie}, \email{stil@ras.ucalgary.ca}, \email{russ@ras.ucalgary.ca}, \email{ricci@ras.ucalgary.ca}, \email{jkgrant@ucalgary.ca}}

\abstract{We present results from simulations of the extragalactic polarized sky at 1.4 GHz. As the basis for our polarization models, we use a semi-empirical simulation of the extragalactic total intensity (Stokes $I$) continuum sky developed at the University of Oxford 
(http://scubed.physics.ox.ac.uk) under the European SKA Design Study (SKADS) initiative, and polarization distributions derived from analysis of polarization observations. By considering a luminosity dependence for the polarization of AGN, we are able to fit the 1.4 GHz polarized source counts derived from the NVSS 
and the DRAO ELAIS N1 deep field survey 
down to approximately 1 mJy. This trend is confirmed by analysis of the polarization of a complete sample of bright AGN. We are unable to fit the additional flattening of the polarized source counts from the deepest observations of the ELAIS N1 survey, which go down to $\sim0.5$ mJy. Below 1 mJy in Stokes $I$ at 1.4 GHz, starforming galaxies become an increasingly important fraction of all radio sources. We use a spiral galaxy integrated polarization model 
to make realistic predictions of the number of polarized sources at $\mu$Jy levels in polarized flux density and hence, realistic predictions of what the next generation radio telescopes such as ASKAP, other SKA pathfinders and the SKA itself will see.}

\FullConference{The 9th European VLBI Network Symposium on The role of VLBI in the Golden Age for Radio Astronomy and EVN Users Meeting\\
                September 23-26, 2008\\
                Bologna, Italy}

\begin{document}

\section*{Simulation Results}
We model the polarized source count data (Figure 1) using a luminosity dependence for the polarization of FRI and FRII AGN described by $\Pi=\Pi_0(L/L_0)^\beta$ with $L_{\rm{0,FRI}}=5.4\times10^{26}$ W~Hz$^{-1}$ and $L_{\rm{0,FRII}}=7.6\times10^{27}$ W Hz$^{-1}$. We use Gaussian initial fractional polarization distributions with $\sigma_{\Pi_0\rm{,FRI}}=\sigma_{\Pi_0\rm{,FRII}}=1.9\%$. We set a maximum allowed $\sigma_{\Pi}$ 
of $5\sigma_{\Pi_0\rm{,FRI}}$ for FRI AGN and $2.5\sigma_{\Pi_0\rm{,FRII}}$ for FRII AGN. Our best fit values of the power-law index $\beta$ are: $\beta_{\rm{FRI}}=-0.5$ and $\beta_{\rm{FRII}}=-0.25$.
From analysis of polarized sources in the NVSS catalogue, \cite{ae} 
found a median fractional polarization ($\Pi_{\rm{med}}$) of 1.77\% for sources with Stokes $I$ fluxes from 100--200 mJy and a $\Pi_{\rm{med}}$ of 1.46\% for sources with $I$ flux greater than 200 mJy. From our model, we find $\Pi_{\rm{med, > 200 mJy}}=1.48\%$ and $\Pi_{\rm{med,100-200 mJy}}=1.84\%$, in good agreement with the observational results from the NVSS.

\begin{figure}[h!]
\begin{center}
\includegraphics[width=12.0cm,clip]{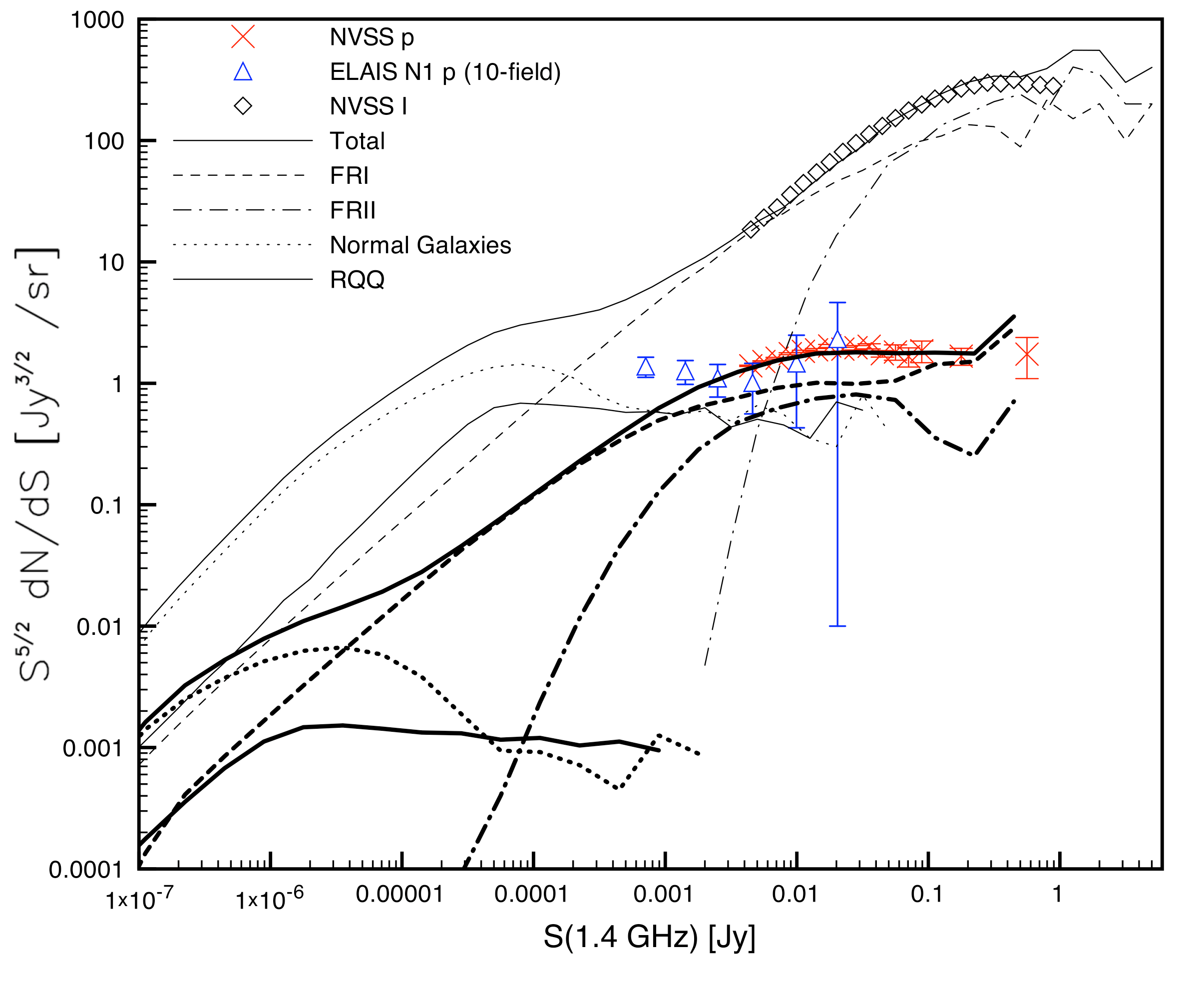}
\caption{Euclidean normalized (S$^{5/2}$ dN/dS [Jy$^{3/2}$ sr$^{-1}$]) total intensity and polarized source counts based on the SKADS Simulated Sky ($S^3$) cosmological simulation \cite{aa}. 
The curves show source counts for FRI (dashed curves), FRII (dot-dashed curves), normal star-forming galaxies (dotted curves), and the solid curves show the sum of the three populations. Radio-quiet quasar (RQQ) source count curves also
included as solid curves. The thin curves show the total intensity source counts and the thick curves show polarized source counts. Total intensity counts from the NVSS are shown as diamonds and polarized counts are shown as crosses. The 10-field DRAO ELAIS N1 polarized source counts \cite{ac} are shown as triangles.}
\label{fig1}
\end{center}
\end{figure}

Ricci et al.~(2009), in preperation, who have redshift information for a complete sample of sources brighter than 1 Jy at 5 GHz, find a correlation between $\Pi_{\rm{1.4 GHz}}$ and $L_{\rm{5 GHz}}$ that directly supports our value of $\beta_{\rm{FRII}}=-0.25$. We are able to find an equally good fit to the polarized source count data using a redshift dependancy for the polarization of FRI and FRII AGN but the data from Ricci et al.~does not favour this interpretation.
However, neither parameterization could provide fits to the additional flattening of the faintest polarized source counts indicating a possible sub-population of highly polarized AGN.
Note that the FRI/FRII distinction, from the $S^3$ simulation~\cite{aa}, 
is not based on morphology but on the traditional FRI/FRII luminosity divide.

Figures 2 is a realization of a $2\times2$ deg$^2$ image of the polarized sky in Stokes Q with a resolution of 18 arcseconds, using the luminosity dependence model for the polarization of FRI and FRII AGN. For the normal star-forming galaxies we use the spiral galaxy model of \cite{ad} 
to generate $\Pi$ distributions at multiple wavelengths. We then interpolate linearly in $\lambda^2$ between the $\Pi$ model values for each galaxy according to it's emitted wavelength (i.e., $\lambda_{\rm{emit}}=\lambda_{\rm{obs}}/(1+z)$). Note that due to the shorter rest frame wavelength, the rise in the polarized source counts for the normal galaxies is steeper than the rise in their Stokes $I$ counts (see Figure 1). A more comprehensive presentation of these results will be included in Stil et al.~(2009), in preparation. 

\begin{figure}[h!]
\includegraphics[width=15.0cm,clip]{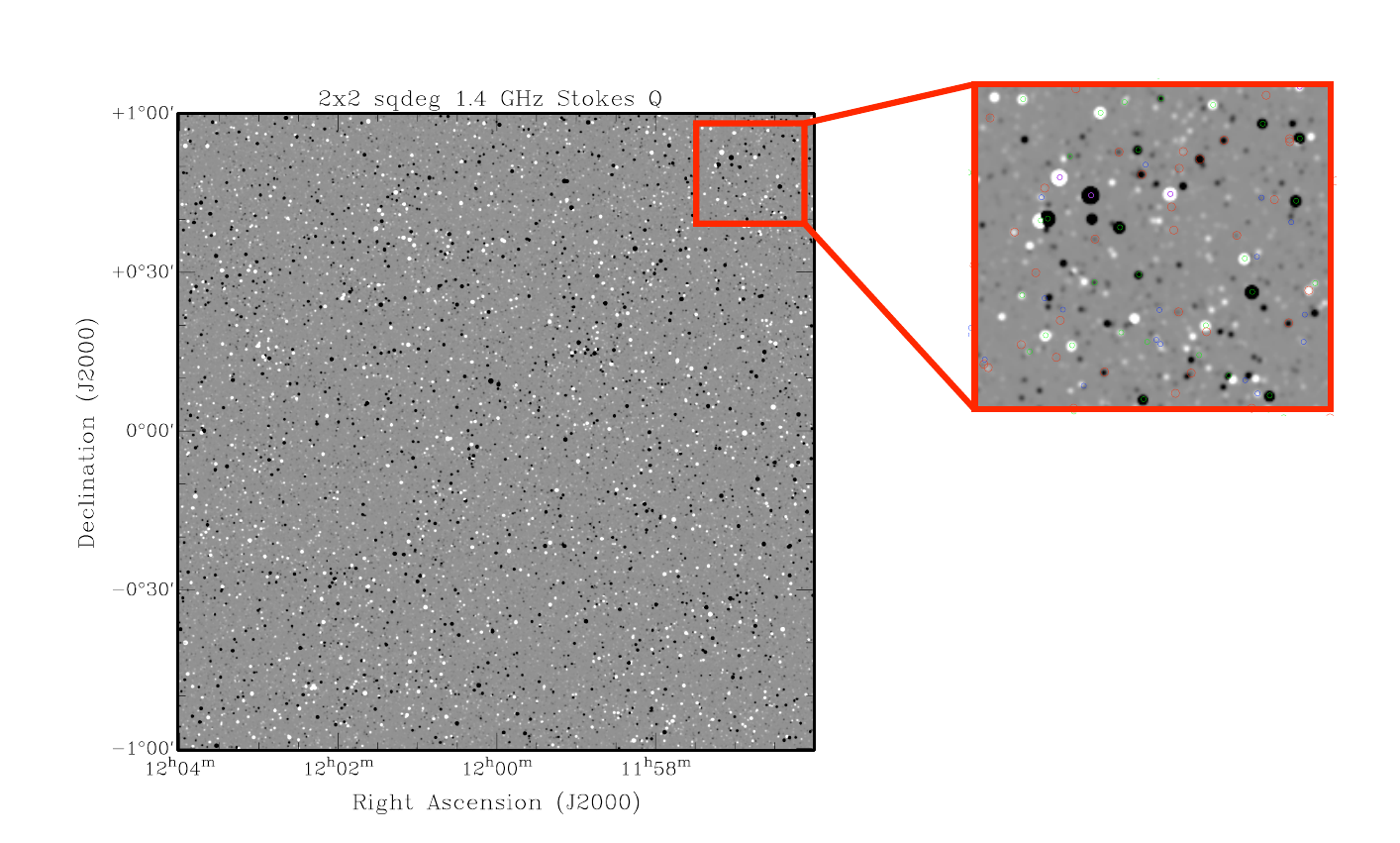}
\caption{$2\times2$ deg$^2$ simulated image of the polarized sky in Stokes Q with a resolution of 18 arcseconds. Image contrast: $\pm0.5$ $\mu$Jy/pixel. Zoomed region with annotations outline different source types with Stokes $I$ flux density greater than 100 $\mu$Jy (red: Normal galaxies, green: FRI, purple: FRII, blue: RQQ). }
\label{fig2}
\end{figure}

\end{document}